\documentclass[12pt,preprint]{aastex}
\usepackage{color}

\slugcomment{Published in ApJ}

\shorttitle{Collapsed Cores in Globular Clusters}
\shortauthors{Djorgovski et al.}

\begin{document}


\title{Study of the diffuse gamma-ray emission from the Galactic plane with ARGO-YBJ}


\author{B.~Bartoli\altaffilmark{1,2},
 P.~Bernardini\altaffilmark{3,4},
 X.J.~Bi\altaffilmark{5},
 P.~Branchini\altaffilmark{6},
 A.~Budano\altaffilmark{6},
 P.~Camarri\altaffilmark{7,8},
 Z.~Cao\altaffilmark{5},
 R.~Cardarelli\altaffilmark{8},
 S.~Catalanotti\altaffilmark{1,2},
 S.Z.~Chen\altaffilmark{5},
 T.L.~Chen\altaffilmark{9},
 P.~Creti\altaffilmark{4},
 S.W.~Cui\altaffilmark{10},
 B.Z.~Dai\altaffilmark{11},
 A.~D'Amone\altaffilmark{3,4},
 Danzengluobu\altaffilmark{9},
 I.~De Mitri\altaffilmark{3,4},
 B.~D'Ettorre Piazzoli\altaffilmark{1,2,21},
 T.~Di Girolamo\altaffilmark{1,2,21},
 G.~Di Sciascio\altaffilmark{8},
 C.F.~Feng\altaffilmark{12},
 Zhaoyang Feng\altaffilmark{5},
 Zhenyong Feng\altaffilmark{13},
 Q.B.~Gou\altaffilmark{5},
 Y.Q.~Guo\altaffilmark{5},
 H.H.~He\altaffilmark{5},
 Haibing Hu\altaffilmark{9},
 Hongbo Hu\altaffilmark{5},
 M.~Iacovacci\altaffilmark{1,2},
 R.~Iuppa\altaffilmark{7,8},
 H.Y.~Jia\altaffilmark{13},
 Labaciren\altaffilmark{9},
 H.J.~Li\altaffilmark{9},
 G.~Liguori\altaffilmark{14,15},
 C.~Liu\altaffilmark{5},
 J.~Liu\altaffilmark{11},
 M.Y.~Liu\altaffilmark{9},
 H.~Lu\altaffilmark{5},
 L.L.~Ma\altaffilmark{5,21}, 
 X.H.~Ma\altaffilmark{5},
 G.~Mancarella\altaffilmark{3,4},
 S.M.~Mari\altaffilmark{6,16},
 G.~Marsella\altaffilmark{3,4},
 D.~Martello\altaffilmark{3,4},
 S.~Mastroianni\altaffilmark{2},
 P.~Montini\altaffilmark{6,16},
 C.C.~Ning\altaffilmark{9},
 M.~Panareo\altaffilmark{3,4},
 L.~Perrone\altaffilmark{3,4},
 P.~Pistilli\altaffilmark{6,16},
 F.~Ruggieri\altaffilmark{6},
 P.~Salvini\altaffilmark{15},
 R.~Santonico\altaffilmark{7,8},
 P.R.~Shen\altaffilmark{5},
 X.D.~Sheng\altaffilmark{5},
 F.~Shi\altaffilmark{5},
 A.~Surdo\altaffilmark{4},
 Y.H.~Tan\altaffilmark{5},
 P.~Vallania\altaffilmark{17,18},
 S.~Vernetto\altaffilmark{17,18},
 C.~Vigorito\altaffilmark{18,19},
 H.~Wang\altaffilmark{5},
 C.Y.~Wu\altaffilmark{5},
 H.R.~Wu\altaffilmark{5},
 L.~Xue\altaffilmark{12},
 Q.Y.~Yang\altaffilmark{11},
 X.C.~Yang\altaffilmark{11},
 Z.G.~Yao\altaffilmark{5},
 A.F.~Yuan\altaffilmark{9},
 M.~Zha\altaffilmark{5},
 H.M.~Zhang\altaffilmark{5},
 L.~Zhang\altaffilmark{11},
 X.Y.~Zhang\altaffilmark{12},
 Y.~Zhang\altaffilmark{5},
 J.~Zhao\altaffilmark{5},
 Zhaxiciren\altaffilmark{9},
 Zhaxisangzhu\altaffilmark{9},
 X.X.~Zhou\altaffilmark{13},
 F.R.~Zhu\altaffilmark{13},
 Q.Q.~Zhu\altaffilmark{5} and
 G.~Zizzi\altaffilmark{20}\\ (The ARGO-YBJ Collaboration)}


 \altaffiltext{1}{Dipartimento di Fisica dell'Universit\`a di Napoli
                  ``Federico II'', Complesso Universitario di Monte
                  Sant'Angelo, via Cinthia, 80126 Napoli, Italy.}
 \altaffiltext{2}{Istituto Nazionale di Fisica Nucleare, Sezione di
                  Napoli, Complesso Universitario di Monte
                  Sant'Angelo, via Cinthia, 80126 Napoli, Italy.}
 \altaffiltext{3}{Dipartimento Matematica e Fisica "Ennio De Giorgi",
                  Universit\`a del Salento,
                  via per Arnesano, 73100 Lecce, Italy.}
 \altaffiltext{4}{Istituto Nazionale di Fisica Nucleare, Sezione di
                  Lecce, via per Arnesano, 73100 Lecce, Italy.}
 \altaffiltext{5}{Key Laboratory of Particle Astrophysics, Institute
                  of High Energy Physics, Chinese Academy of Sciences,
                  P.O. Box 918, 100049 Beijing, P.R. China.}
 \altaffiltext{6}{Istituto Nazionale di Fisica Nucleare, Sezione di
                  Roma Tre, via della Vasca Navale 84, 00146 Roma, Italy.}
 \altaffiltext{7}{Dipartimento di Fisica dell'Universit\`a di Roma
                  ``Tor Vergata'', via della Ricerca Scientifica 1,
                  00133 Roma, Italy.}
 \altaffiltext{8}{Istituto Nazionale di Fisica Nucleare, Sezione di
                  Roma Tor Vergata, via della Ricerca Scientifica 1,
                  00133 Roma, Italy.}
 \altaffiltext{9}{Tibet University, 850000 Lhasa, Xizang, P.R. China.}
 \altaffiltext{10}{Hebei Normal University, Shijiazhuang 050016,
                   Hebei, P.R. China.}
 \altaffiltext{11}{Yunnan University, 2 North Cuihu Rd., 650091 Kunming,
                   Yunnan, P.R. China.}
 \altaffiltext{12}{Shandong University, 250100 Jinan, Shandong, P.R. China.}
 \altaffiltext{13}{Southwest Jiaotong University, 610031 Chengdu,
                   Sichuan, P.R. China.}
 \altaffiltext{14}{Dipartimento di Fisica dell'Universit\`a di
                   Pavia, via Bassi 6, 27100 Pavia, Italy.}
 \altaffiltext{15}{Istituto Nazionale di Fisica Nucleare, Sezione di Pavia,
                   via Bassi 6, 27100 Pavia, Italy.}
 \altaffiltext{16}{Dipartimento di Fisica dell'Universit\`a ``Roma Tre'',
                   via della Vasca Navale 84, 00146 Roma, Italy.}
 \altaffiltext{17}{Osservatorio Astrofisico di Torino dell'Istituto Nazionale
                   di Astrofisica, via P. Giuria 1, 10125 Torino, Italy.}
 \altaffiltext{18}{Istituto Nazionale di Fisica Nucleare,
                   Sezione di Torino, via P. Giuria 1, 10125 Torino, Italy.}
 \altaffiltext{19}{Dipartimento di Fisica dell'Universit\`a di
                   Torino, via P. Giuria 1, 10125 Torino, Italy.}
 \altaffiltext{20}{Istituto Nazionale di Fisica Nucleare - CNAF, Viale
                   Berti-Pichat 6/2, 40127 Bologna, Italy.}
 \altaffiltext{21}{Corresponding authors: llma@ihep.ac.cn, dettorre@na.infn.it, digirola@na.infn.it}

\begin{abstract}

The events recorded by ARGO-YBJ in more than five years of data collection have been
analyzed to determine the diffuse gamma-ray emission in the Galactic plane
at Galactic longitudes $25^\circ <l<100^\circ$ and Galactic latitudes
$|b|<5^\circ$. The energy range covered by this analysis, from $\sim$350 GeV
to $\sim$2 TeV, allows the connection of the region explored by \emph{Fermi} with the
multi-TeV measurements carried out by Milagro. Our analysis has been focused
on two selected regions of the Galactic plane, i.e.,
$40^\circ <l<100^\circ$ and  $65^\circ <l<85^\circ$ (the Cygnus region), where
Milagro observed an excess with respect to the predictions of current models.
Great care has been taken in order to mask the most intense gamma-ray sources,
including the TeV counterpart of the Cygnus cocoon recently identified by
ARGO-YBJ, and to remove residual contributions. The ARGO-YBJ results do not
show any excess at sub-TeV energies corresponding to the excess found by
Milagro, and are consistent with the predictions of the \emph{Fermi} model
for the diffuse Galactic emission. From the measured energy distribution we
derive spectral indices and the differential flux at 1 TeV of the diffuse
gamma-ray emission in the sky regions investigated.

\end{abstract}


\section{Introduction}

Diffuse gamma-rays are the sum of contributions from several
components: the truly diffuse Galactic gamma-rays produced by the
interaction of cosmic rays, the extragalactic background and the
contribution from undetected and faint Galactic gamma-ray sources. On the
Galactic plane the Galactic gamma-rays dominate the other components. The
processes leading to this emission are the interaction of cosmic nuclei
with the interstellar gas through the production and decay of secondary
$\pi^0$ mesons, the bremsstrahlung of high-energy cosmic electrons, and
their inverse Compton scattering on low-energy interstellar radiation
fields. The spectrum of this radiation may provide insight into the
propagation and confinement in the Galaxy of the parent cosmic rays, their
source distribution and their spectrum at the acceleration sites. Since
gamma-rays are not deflected by magnetic fields, the diffuse component traces
cosmic rays and the interstellar environment in distant parts of the
Galaxy. In addition, the Galactic diffuse emission represents the natural
background to many different signals \citep{Moskalenko2004}. The
knowledge of the diffuse emission is necessary for the accurate detection
of gamma-ray sources, either as point-like or extended. High-quality data
on the diffuse emission in the Galactic Center region are needed to
constrain the dark matter models \citep{Macias}.
Electrons and positrons from astrophysical
sources invoked to explain the PAMELA results also induce gamma-rays when
propagating in the Galaxy. The induced gamma-rays represent an
additional contribution to the diffuse gamma-ray background\citep{Zhang2010}.

Galactic diffuse gamma-rays with energies from 100 MeV to a few GeV were
firstly detected by the space-borne detectors SAS-2 \citep{SAS2}
and COS-B \citep{COSB}, which revealed
the noticeable correlation between the flux of gamma-rays and the density
of the interstellar medium. Then COMPTEL and EGRET on board the \emph{Compton
Gamma Ray Observatory} provided the first maps unveiling the
spectrum of the diffuse emission from 1 MeV up to
10 GeV \citep{Hunter1997,Strong2011}.
In the GeV energy range, the EGRET data show in all directions a
significant excess (the so-called ``GeV excess'') with respect to the
predictions obtained assuming a cosmic-ray flux as that measured at Earth.
This result stimulated intense theoretical studies and many possible
explanations have been proposed, including harder cosmic-ray spectra
throughout the Galaxy \citep{Strong2000}, large contribution
from high-energy electrons via inverse Compton scattering \citep{Porter},
or production of gamma-rays from annihilation of dark matter particles with
mass 40$-$50 GeV\citep{Bi2008}. At intermediate
Galactic latitudes this excess has not been confirmed by the Large Area
Telescope (LAT) on board the\emph{ Fermi Gamma Ray Space Telescope} launched
in 2008 \citep{abdo2009}. With a
sensitivity more than one order of magnitude better than EGRET, \emph{Fermi}-LAT
mapped the gamma-ray sky up to a few hundreds of GeV with
unprecedented accuracy. Observation and discovery of gamma-ray discrete
sources \citep{Nolan, Ackermann2013, Acero, abdo2013} as well as the study
of the GeV diffuse emission at different Galactic latitudes
\citep{Ackermann2012a} have been the primary targets of these analyzes.
At higher energies, due to their low fluxes, the diffuse gamma-rays can be
efficiently studied only by ground-based detectors with large effective
areas. The Imaging Atmospheric Cherenkov Telescopes (IACTs),
such as Whipple and HEGRA, have set upper limits in very
narrow regions \citep{LeBohec2000, Aharonian2001}. The H.E.S.S. telescope
array has carried out a survey of the Galactic plane covering the region
$-75^\circ <l<60^\circ$ in longitude at an energy threshold of about 250 GeV,
and has presented the latitude profile
of diffuse emission at $|b|<2^\circ$ \citep{Abramowski}.
This represents the first observational assessment of diffuse TeV
gamma-rays by a Cherenkov telescope. The spatial correlation of gamma-rays
with giant molecular clouds in the inner Galaxy has been also
investigated, the hardness of gamma-ray spectrum indicating that these
gamma-rays originate from protons and nuclei rather than
electrons \citep{Aharonian2006a}. However, though IACTs are the most
sensitive detectors operating in the field of gamma-ray astronomy, limited
by their field of view (FOV) they are not \textbf{well suited}
to observe diffuse
gamma-rays that have a large-scale structure. On the other hand, air shower
arrays providing a large FOV and a very high duty cycle look more adequate for
sky survey purposes. Many air
shower experiments have set upper limits to diffuse gamma-rays, such as
Tibet AS$\gamma$ at TeV energies \citep{Amenomori2006}, and EAS-TOP
\citep{Aglietta}, KASCADE \citep{KASCADE}, and
CASA-MIA \citep{CASAMIA} at energies $>$100 TeV.
The Milagro detector has made the first positive observation of the
diffuse gamma-ray flux from the Galactic plane, measuring the integral
flux above 3.5 TeV in the region $40^\circ <l<100^\circ$, $|b|<5^\circ$
\citep{Atkins05}. Assuming that the
contribution from inverse Compton scattering is negligible at TeV energies,
the Milagro measurement is many times higher than expected, indicating the
existence of a ``TeV excess'' possibly connected to the ``GeV excess'' of
diffuse gamma-rays observed by EGRET \citep{TeVexcess}.
Most notably, in a following analysis of seven-years data, the Milagro
collaboration reported a clear excess of diffuse gamma-rays at 15 TeV
median energy from the galactic region $65^\circ <l<85^\circ$,
suggesting the presence of active cosmic-ray sources
accelerating hadrons \citep{abdo2008}. This interpretation is
supported by the fact that this longitude interval harbors the Cygnus X
star-forming region, rich with possible cosmic-ray acceleration sites.
However, the \emph{Fermi}-LAT data do not confirm at GeV energies the broadly
distributed excess of diffuse emission observed by Milagro at multi-TeV
energies \citep{Ackermann2012b}. Instead, they have found a bright
extended gamma-ray source centered at $l=79^{\circ}.6\pm 0^{\circ}.3$,
the so-called ``Cygnus cocoon'' \citep{Ackermann2011},
whose TeV counterpart has been recently identified by the ARGO-YBJ experiment
\citep{Bartoli2014a}. These results suggest that the gamma-ray excess of the
cocoon is likely due to a population of freshly accelerated
hadronic cosmic rays, even
though a leptonic or mixed origin cannot be discarded.
High-energy protons fastly
diffusing from this source and interacting with the local gas could
generate gamma-rays that contribute to the diffuse excess observed by
Milagro. To better clarify the interpretation of these results, we have
used the data collected by ARGO-YBJ in 5.3 yr to measure
the diffuse gamma-ray emission in the $25^\circ <l<100^\circ$, $|b|<5^\circ$
Galactic region, then selecting the two regions $40^\circ <l<100^\circ$
and $65^\circ <l<85^\circ$ in order to compare the results with the Milagro
measurements. Thanks to its high-altitude location and its particular layout,
ARGO-YBJ is able to operate at an energy threshold of about 300 GeV,
in such a way providing results that bridge the \emph{Fermi} GeV energies and the
multi-TeV range explored by Milagro.

The plan of the paper is as follows. In Section 2 we describe the
ARGO-YBJ detector and its performance. In Section 3 we present data
selection and data analysis methods. The results of the analysis are
presented in  Section 4. Section 5 is devoted to a summary of the results
and to the conclusions.

\section{The ARGO-YBJ experiment}

The ARGO-YBJ detector, hosted in a building at the YangBaJing Cosmic Ray
Observatory (Tibet, China, 90$^\circ$31'50" E, 30$^\circ$06'38" N),
4300 m above sea level, has been designed for very high-energy (VHE) gamma-ray
astronomy and cosmic-ray observations. It is made up of a single layer of
resistive plate chambers (RPCs) operated in streamer mode, 2.850 m $\times$
1.225 m each, organized in a modular configuration to cover a
surface of about 5600 m$^2$ with an active area of about 93\%.
The RPCs detect the charged particles in air showers
with an efficiency $\ge 98\%$.
To improve the shower reconstruction,
other chambers are deployed around the central carpet for a total
instrumented area of 100 m $\times$ 110 m.
A highly segmented readout is
performed by means of 55.6 cm $\times$ 61.8 cm external electrodes,
called "pads," whose fast signals are used for triggering and timing purposes.
These pads provide the digital readout of the detector up to 22
particles/m$^2$, allowing the count of the air shower charged particles
without any significant saturation up to primary cosmic-ray energies of
about 200 TeV \citep{Bartoli2012a}.
In order to extend the dynamical range to PeV
energies each RPC is also equipped with two large size pads
(139 cm $\times$ 123 cm) allowing the collection of the total charge developed
by the particles hitting the detector \citep{aielli12}. The
digital output of each pad is splitted in two signals sent to the logic
chain that builds the trigger and to the 18,360 multi-hit time-to-digital
converters, which are routinely calibrated with 0.4 ns accuracy by means of
an off-line method using cosmic-ray showers \citep{he07, aielli09a}. More
details about the detector and the RPC performance can be found in
\citet{aielli06} and \citet{aielli09b}.
The detector is connected to two independent acquisition systems
corresponding to two different operation modes, referred to as the shower
mode and the scaler mode \citep{aielli08}. The data used in this
paper were recorded by the digital readout in shower mode. This mode is
implemented by means of an inclusive trigger based on the time correlation
between the pad signals, depending on their relative distance. In this way
the data acquisition is triggered when at least 20 pads in the central
carpet are fired in a time window of 420 ns. By means of this trigger the
energy threshold for gamma-induced showers can go down to 300 GeV with an
effective area depending on the zenith angle
(see Figure 1 in \citet{Bartoli2013}).

The event reconstruction follows a standard procedure allowing a detailed
space-time reconstruction of the shower front, including the position of
the shower core and the incident direction of the primary particle.
A detailed account of the reconstruction algorithm can be found in
\citet{Bartoli2011a, Bartoli2011b, Bartoli2013, Bartoli2015}.
Briefly, the shower core position is obtained by fitting the lateral density distribution of the shower charged particles with a Nishimura-Kamata-Greisen-like function using the maximum likelihood method.
The arrival direction of the showers is reconstructed by the least squares method assuming a conical shape for the shower front, as described by Equation (1) in \citep{aielli09a}, which gives the relation between the particle arrival
time and the distance to the shower core.
The angular resolution depends on the number of fired pads \emph{N}$_{pad}$.
The opening angle $\psi_{70}$ containing 71.5\% of the events from a point
source is about 2$^\circ$ for events with \emph{N}$_{pad} > 20$,
1.36$^\circ$ for \emph{N}$_{pad} > 60$ and 0.99$^\circ$ for events with
\emph{N}$_{pad} > 100$.
The number of hit pads \emph{N}$_{pad}$ is the observable related to the
primary energy. However, the number of particles at ground level is not a
very accurate estimator of the primary energy of the single event, due to the
large fluctuations in the shower development in the atmosphere and to its
partial sampling with the limited detector area.
The primary energy distribution corresponding to
different \emph{N}$_{pad}$ intervals is very broad, spanning over more than one
order of magnitude for small \emph{N}$_{pad}$ values. The relation between \emph{N}$_{pad}$
and the primary gamma-ray energy of the selected showers is reported in
\citet{Bartoli2015}. Since the variable \emph{N}$_{pad}$ does not allow the
accurate measurement of the primary energy of the single event, the energy
spectrum is evaluated by studying the global distribution of \emph{N}$_{pad}$. The
observed distribution is compared with a set of simulated ones obtained with
different test spectra, in order to find out that which better
reproduces the data (as carried out in Section 4.1).
The angular resolution, pointing
accuracy, absolute energy calibration and detector stability are
tested by measuring the cosmic-ray shadow cast by the Moon, detected with
a significance of 10 standard deviations (s.d.) per month \citep{Bartoli2011b}.

\section{Data analysis}

The ARGO-YBJ experiment began taking data in its full configuration in
 2007 November at a trigger rate of 3.5 kHz with a dead time of 4\%.
It has been operated stably for more than five years,
up to 2013 January, with an average duty cycle of 86\%, for a total
effective time of 1670.45 days. For the present analysis, events with
zenith angles less than 50$^\circ$, corresponding to the declination interval
$-20^\circ < \delta <80^\circ$, and \emph{N}$_{pad} >20$ are used. A set of standard
cuts, applied to the shower core reconstructed position and to the time
spread of the shower front, have been used to select high-quality data.
With this data selection a total of $6.407\times 10^{10}$  shower
events are observed
from the Galactic plane in the latitude belt $|b|<15^\circ$.
The fraction of survived events is about 80\%.
With these selections more background cosmic rays than gamma-rays are rejected,
implying an increase of the sensitivity \citep{Bartoli2013}.
These data have been used to measure the diffuse emission from the
regions of the Galactic plane of longitude $25^\circ <l<100^\circ$ and
$130^\circ <l<200^\circ$. Indeed the region $100^\circ <l<130^\circ$ is
excluded since in the high declination region $\delta > 60^\circ$ the Galactic
plane runs parallel to the right ascension axis and the contribution from the
signal could affect the background estimation.
The excess of gamma-induced showers is obtained following the
procedure of the background estimation applied to the ARGO-YBJ
data as reported in \citet{Bartoli2011a, Bartoli2013}.
All data are divided into three pad groups, $20 <N_{pad} \le 59$,
$60 \le N_{pad} \le 99$ and $N_{pad} \ge 100$.
For each group of pad multiplicity both
the inner and the outer regions of the Galactic plane are divided into a
grid of $0.1^\circ \times 0.1^\circ$ bins and filled with the detected events
according to their reconstructed arrival directions (event map).
The number of cosmic-ray background events (background map) is estimated by
using the direct integration method of
\citet{fleysher04}. The effect of cosmic-ray anisotropy on the background
evaluation has been estimated and corrected by applying the normalization
given in \citet{Bartoli2011a}. This procedure is applied to each map bin
using a surrounding region of $16^\circ \times 16^\circ$ in which the
estimated background is renormalized to the detected events.
The $\pm 5^\circ$ region
around the Galactic plane and the $4^\circ \times 4^\circ /cos(b)$ region
around the Crab Nebula position are excluded from the normalization procedure.
However, since the diffuse gamma-ray emission extends to more than
$|b|=5^\circ$, its contribution causes an overestimation of the correction
related to the cosmic-ray anisotropy. This effect has been evaluated
using the latitude profile provided by the \emph{Fermi}-LAT model for the diffuse
Galactic emission (see section 4) smeared out with the ARGO-YBJ point spread
function (PSF). A variation of 15\% of this contribution
implies on average a variation of the excess of about 4\%.
To investigate the systematic errors related to the extension of the region
used to evaluate the anisotropy effect, the box size has been varied from
$12^\circ \times 12^\circ$ to $20^\circ \times 20^\circ$
obtaining an excess variation of $\sim$10\%. Both the event and
background maps have been smoothed with the
PSF corresponding to each \emph{N}$_{pad}$ interval. Then the background map has been
subtracted to the event map obtaining the event excess map. This map contains
events from true diffuse gamma-rays as well as from point and extended
sources, whereas, due to the background subtraction, the isotropic
extragalactic emission is canceled out. Indeed, many TeV gamma-ray sources
of different extension lying on or close to the Galactic plane have been
detected in the longitude range $25^\circ <l<100^\circ$. Source locations as
given in the TeVCat\footnote{http://tevcat.uchicago.edu} are excluded from the
analysis. Faint sources (SNR G54.1+0.3, VER J2016+372, HESS J1923+141 (W51)
and HESS J1943+213) have not been masked. Their total contribution to the
diffuse flux at 1 TeV is estimated to be about 2.5\%.

Taking into account the angular resolution of the detector and the
extension of these sources, the contribution from a region
$4^\circ \times 4^\circ /cos(b)$
centered around each source location has been removed.
Some boxes include two sources, with the fainter one near its edge (as for
instance HESS J1849-000 and VER J2019+368).
Sources distant less than 1$^{\circ}$.2, as for instance HESS J1857+026 and HESS
J1858+020, have been masked with a unique box centered at the median point.
The massive star-forming region of Cygnus X hosts the extended cocoon,
firstly observed by \emph{Fermi-LAT} above 1 GeV \citep{Ackermann2011},
whose emission at TeV energies has been recently
assessed by ARGO-YBJ \citep{Bartoli2014a}.
Due to its extension of about 2$^\circ$, this region has been masked with a box
$6^\circ \times 6^\circ /cos(b)$ centered on the source position found by
ARGO-YBJ. The chosen dimensions of these boxes is a compromise between a
desired large excluded region, in order to minimize the contamination from the
sources, and the requirement of not reducing the statistics.
With this choice the solid angle of the region $25^\circ <l<100^\circ$,
$|b|<5^\circ$ is reduced of about 22\%.
The spillover from these sources outside the masked regions has been estimated
by tracking their path inside the FOV of ARGO-YBJ.
The contamination is calculated bin by bin and subtracted from the total
excess in the $0^{\circ}.1 \times 0^{\circ}.1$ bin event excess map.
For ARGO J1839-0627/HESS J1841-055, ARGO J1907+0627/MGRO J1908+06, and
ARGO J2031+4157 (the Cygnus cocoon) the fluxes measured by ARGO-YBJ have been
considered \citep{Bartoli2013, Bartoli2014a}.
Since the PSF broadens with decreasing energy, this contamination is found
higher for the first energy bin (corresponding to the group with
$20< N_{pad} \le 59$), with an average value of 14\%, while it is 21\% in
the Cygnus region $65^\circ < l < 85^\circ$.

\section{Results}

\begin{figure}[!t]
\vspace{5mm} \centering \epsscale{.8}
\plotone{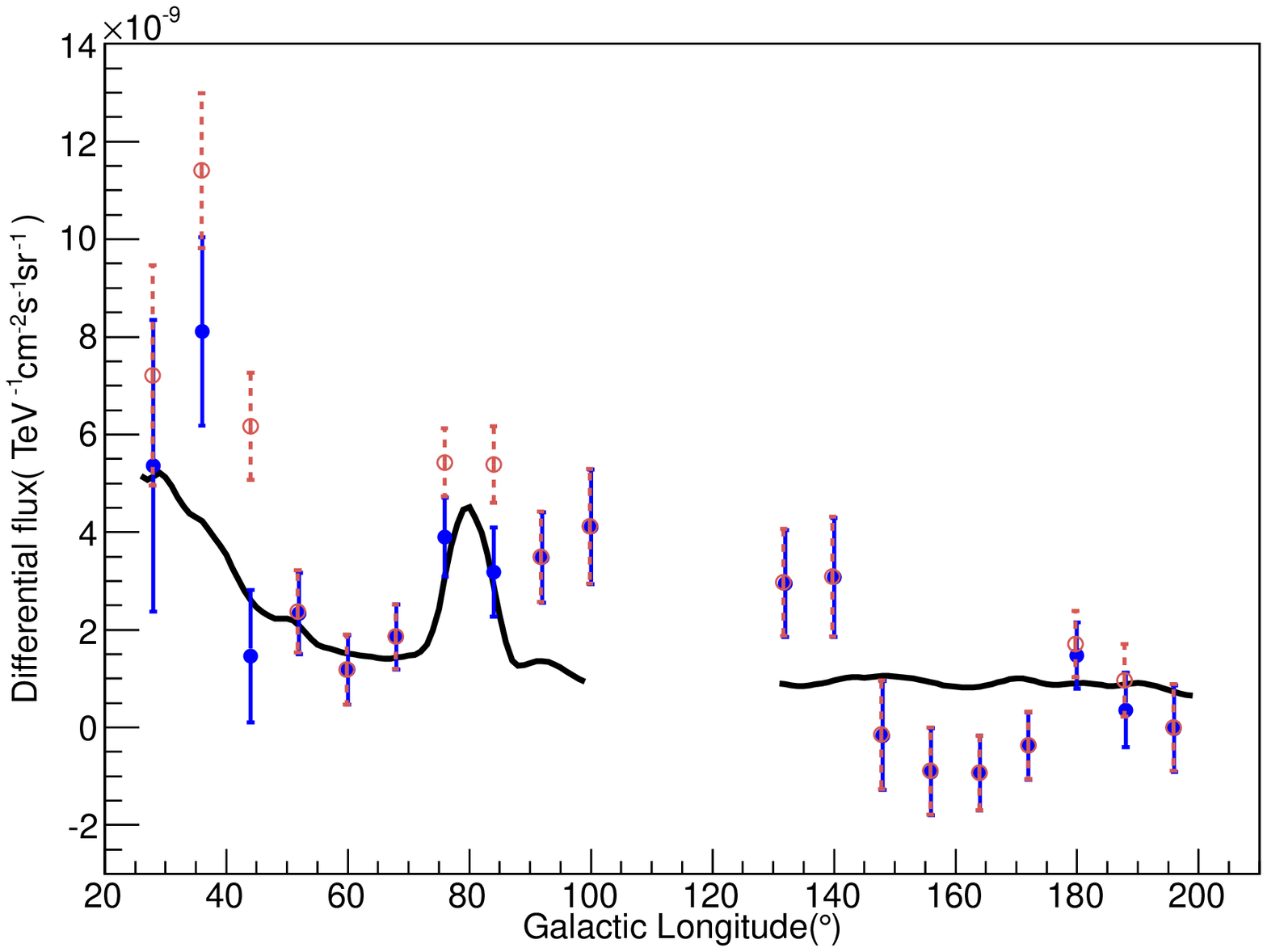}
\caption{Galactic longitude profile of the diffuse gamma-ray emission in the
Galactic latitude interval $|b|<5^\circ$ at an energy of 600 GeV as
obtained from the ARGO-YBJ data. The filled circles show the results
after masking the sources, while the open circles show the results without the
masking. The solid line represents the value quoted by
the \emph{Fermi}-DGE model at the same energy, smeared out with the ARGO-YBJ PSF.}
\label{longitude}
\end{figure}

\begin{figure}[!t]
\vspace{5mm} \centering \epsscale{.8}
\plotone{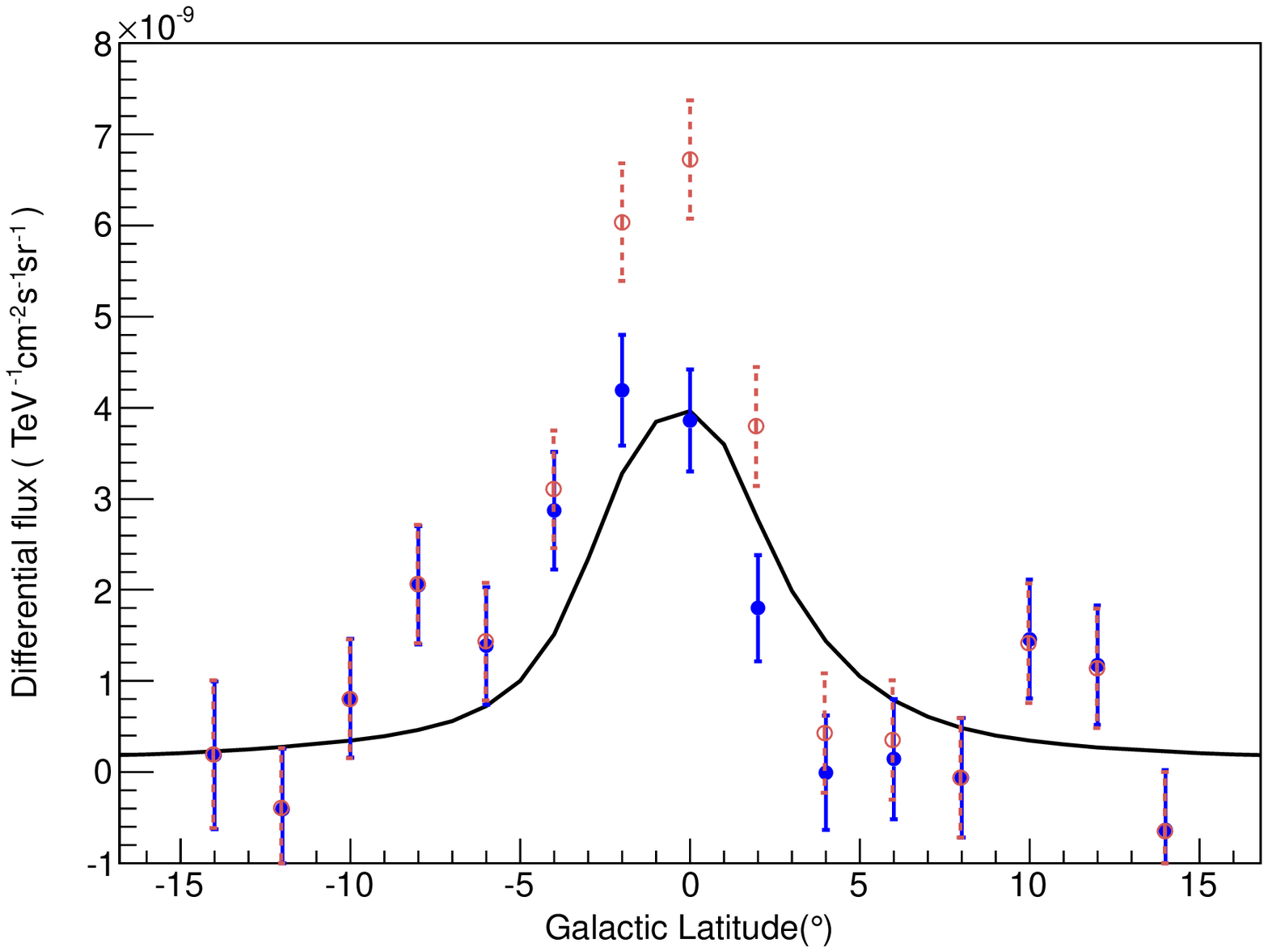}
\caption{Galactic latitude profile of the diffuse gamma-ray
emission in the Galactic longitude interval $25^{\circ}<l<100^{\circ}$ at an
 energy of 600 GeV as obtained from the ARGO-YBJ data. The filled
circles show the results after masking the sources, while the open circles
show the results without the masking. The solid line
represents the value quoted by the \emph{Fermi}-DGE model at the same energy, smeared
out with the ARGO-YBJ PSF.}
\label{latitude}
\end{figure}

The Galactic longitude profile of the diffuse gamma-ray emission at 600
GeV in the latitude belt $|b| < 5^\circ$ is shown in Fig. \ref{longitude}
by the filled circles. The profile obtained without masking the
sources is also plotted as the open circles.
Each point represents the flux obtained with the spectral analysis reported
in Section 4.1 and averaged over 8$^{\circ}$ longitude bins.
The flux is evaluated taking into account all the events with \emph{N}$_{pad} >20$.
For each group of pad multiplicity the event excess measured in each longitude
bin is converted to a flux using the effective areas estimated by means of a
full Monte Carlo simulation of extensive air showers \citep{CORSIKA}
and of the RPC array \citep{G4argo}. Then a spectral analysis is
carried out as described in Section 4.1. The negative fluxes which appear in
the outer Galaxy profile correspond to negative excess values in the
event excess map. In this case a spectral index -2.7 has been assumed.
The significance of the excess measured in the $25^{\circ}< l < 100^{\circ}$
region is 6.9 s.d., while no  excess is detected in the outer Galaxy region
$130^{\circ} < l < 200^{\circ}$. Fig. \ref{latitude} shows for
$|b|< 15^{\circ}$ the Galactic latitude profile of the excess in bins of
2$^{\circ}$. The filled circles show the results after masking the
sources, while the open circles show the results without masking.
The continuous line in these plots represents the flux at 600
GeV  provided by the standard \emph{Fermi}-LAT model for the  diffuse Galactic
emission $gal\_2yearp7v6\_v0.fits$ (hereafter \emph{Fermi}-DGE) to which we refer
for comparison with the ARGO-YBJ data. This model, available at the Fermi
Science Support Center\footnote{http://fermi.gsfc.nasa.gov/ssc/data/access/lat/BackgroundModels.html}, has been used to generate four \emph{Fermi}-LAT
Source Catalogs \citep{Nolan, Ackermann2013, Acero, abdo2013}.
For the first time a diffuse flux measured by a ground-based detector overlaps
and can be compared with results from direct measurements. In the region
$25^{\circ}< l < 100^{\circ}$ we find a satisfactory general agreement between
the ARGO-YBJ data and the fluxes predicted by \emph{Fermi}-DGE, mostly in the inner
range $40^{\circ}< l < 90^{\circ}$. The maximum deviations $\leq$2.5 s.d. are
observed at three values outside this interval. In addition to statistical
fluctuations, systematic uncertainties related to the background evaluation,
imperfect modeling of the Galactic diffuse emission and other effects as,
for instance, an energy-dependent diffuse flux from unresolved sources, can
contribute to this discrepancy. The amount of such uncertainties is addressed
in the next section, where the results of a spectral analysis of the flux
detected in this region are presented. In the following sections, the spectral
analyses concerning two selected subregions, $40^{\circ}< l < 100^{\circ}$
and $65^{\circ}< l < 85^{\circ}$, and the upper limit to the diffuse flux in
the outer Galaxy are reported and discussed.

\subsection{The Galaxy Region $25^{\circ} < l < 100^{\circ}$,
$|b| < 5^{\circ}$}

\begin{figure}[!t]
\vspace{5mm} \centering \epsscale{.8}
\plotone{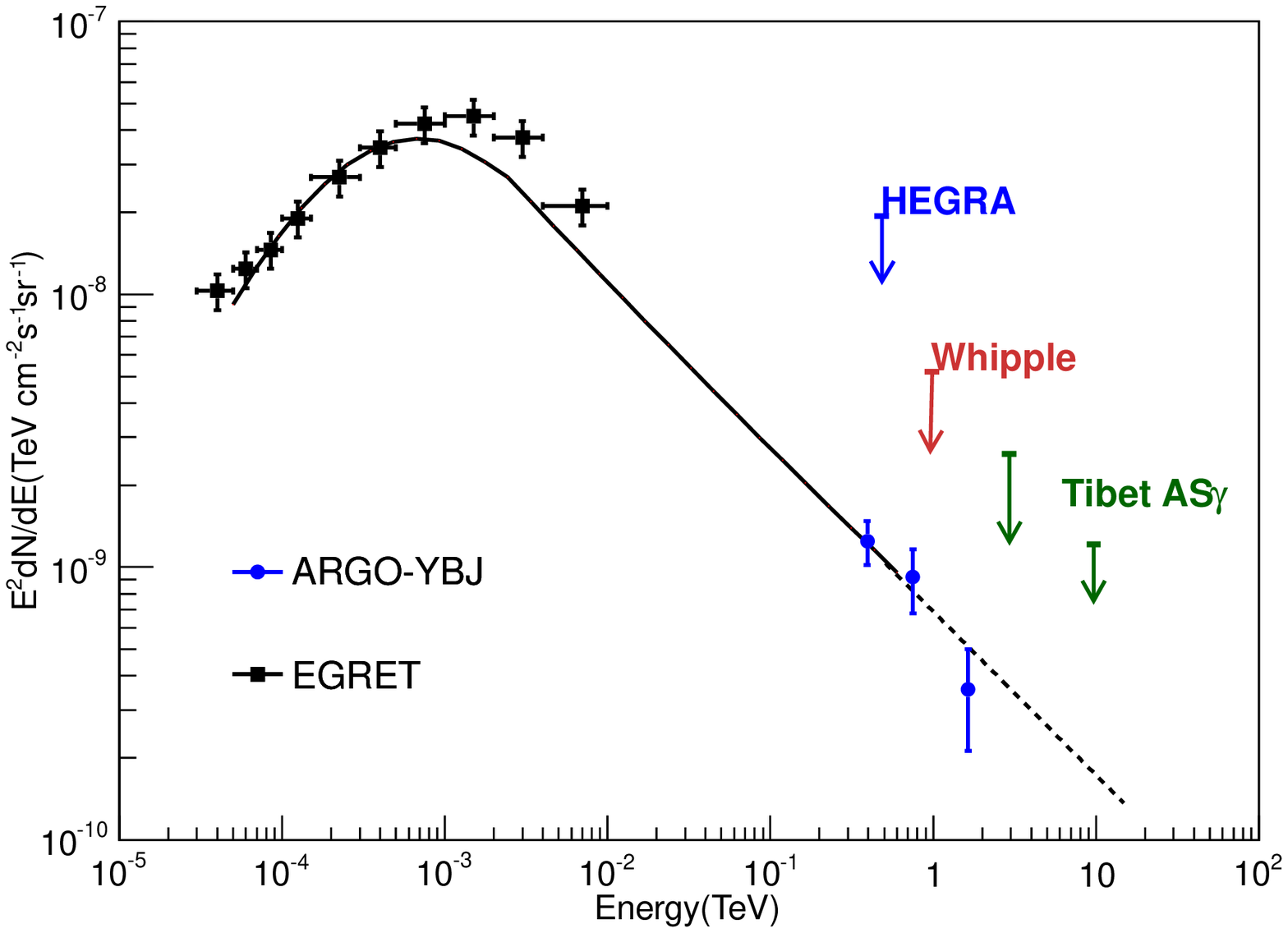}
\caption{The energy spectrum of the diffuse gamma-ray emission measured by
ARGO-YBJ in the Galactic region $25^{\circ}<l<100^{\circ}$, $|b|<5^{\circ}$
(dots). The solid line shows the flux in the same region according to the
\emph{Fermi}-DGE model. The short-dashed line represents its extension following a
power law with spectral index -2.6. The EGRET results (squares) in the same
Galactic region $25^{\circ}<l<100^{\circ}$, $|b|<5^{\circ}$ and the
upper limits quoted by HEGRA (99\% C.L.,
$38^{\circ}<l<43^{\circ}$, $|b|<2^{\circ}$), Whipple (99.9\% C.L.,
$38.5^{\circ}<l<41.5^{\circ}$, $|b|<2^{\circ}$) and Tibet AS$\gamma$
(99\%C.L., $20^{\circ}<l<55^{\circ}$, $|b|<2^{\circ}$) are also shown.}
\label{25_100}
\end{figure}

The total number of shower events recorded in this region is
$7.92\times 10^9$. To carry out a spectral analysis  with a distribution of the
number of events in excess as a function of \emph{N}$_{pad}$ we follow the method
described in \citet{aielli2010}. Sampling events are generated by simulations
in the energy range from 10 GeV to 100 TeV assuming the spectral index of a
power law as a parameter and taking into account the detailed ARGO-YBJ detector
response. The fit to the data is made by comparing the measured excess
in each pad multiplicity interval
with simulations. A differential spectral index -2.80$\pm$0.26 is found.
The corresponding median energies of the events recorded in the \emph{N}$_{pad}$
intervals $20 < N_{pad} \le 59$, $60 \le N_{pad} \le 99$, $N_{pad} \ge 100$
are 390 GeV, 750 GeV, and 1.64 TeV, respectively.
Since the median energies depend on the spectral index, these values are
affected by an uncertainty of about 30\%.
This result is shown in Fig. \ref{25_100} (dots).
Upper limits from HEGRA, Whipple, and
Tibet AS$\gamma$ experiments are also shown. The solid line
represents the expectation according to the \emph{Fermi}-DGE model. This model is
defined between 50 MeV and 600 GeV. Above 10 GeV the spectrum follows a
power law with spectral index about -2.6 and has been extended with the same
slope to  TeV energies as a guide for the eye (dashed line).
The \emph{Fermi}-DGE model is based on
template fits to the all sky gamma-ray data and includes an Inverse Compton
component generated by the GALPROP\textbf{\footnote{http://sourceforge.net/projects/galprop}}
cosmic-ray propagation code \citep{Strong2000, Vlad}.
It arises from an accurate comparison
of data to the sum of many contributions including the ones from detected
sources. It is not easy to assess the uncertainty associated to the
predictions of the model.
In \citet{Ackermann2012a} a grid of models is considered, which
represents well the gamma-ray sky with an agreement within 15\% of the
data, although various residuals at a $\sim$30\% level are found, both at
small and large scales. Similar systematic errors
are quoted in \citet{Macias} and \citet{Gordon}
in an analysis concerning a very narrow region around the Galactic center.
Above 40 GeV the accuracy of the modeling is limited primarily by the photon
statistics and the diffuse emission has been derived by
extrapolating the emissivities measured at lower energies.
The average systematic uncertainty at high energies is expected to be
greater than 10\%, which is the systematics affecting the determination of
the effective area \citep{Ackermann2012c}.
Two main systematic uncertainties can affect the ARGO-YBJ flux estimate:
one on the background and the other on the absolute scale energy.
The systematics on the background evaluation has been discussed in Section 3.
The energy scale reliability has been checked by studying the westward
shift of the cosmic-ray shadow cast by the Moon due to the geomagnetic field.
At TeV energies the total absolute energy scale error is less than 13\%
\citep{Bartoli2011b} and the corresponding systematic error on the flux
normalization would be about 23\%. Minor contributions to the
systematic error come from the uncertainty on the residual contamination
of the masked sources, the ARGO-YBJ and H.E.S.S. fluxes being quoted with
about 30\% systematic uncertainty \citep{Bartoli2012b, Aharonian2006b},
from the uncertainty ($<$4\%) on the detector efficiency and from the
systematic error of about 5\% affecting the effective area estimate.
We combine these various errors in quadrature to obtain a total systematic
error of $\sim$27\%.
The estimated ARGO-YBJ flux at 1 TeV is
$(6.0\pm 1.3)\times 10^{-10}$ TeV$^{-1}$ cm$^{-2}$ s$^{-1}$ sr$^{-1}$,
13\% lower than the prediction based on the \emph{Fermi}-DGE extrapolation.
Taking into account the whole uncertainties, we deem the ARGO-YBJ data set
consistent with the model predictions.

\subsubsection{Treatment of the unresolved sources}

However, it is worth noting that
part of the detected signal could originate in faint sources that are
unresolved because of their low flux. Indeed, the Galactic region that we are
studying hosts many potential gamma-ray sources, mainly supernova remnants
(SNRs) and pulsar wind nebulae (PWNe). Shell-type SNRs and SNRs interacting
with molecular clouds form an established source class in VHE gamma-ray
astronomy \citep{TeVreview}, although whether the nature of their
emission is predominantly hadronic or leptonic is still a matter of debate
\citep{Yuan, Volk}. The TeV flux depends on
the energy available for shock acceleration, on the distance and on many
environmental parameters. The majority of the identified Galactic TeV
sources are PWNe \citep{deOna}. Ninety Galactic PWNe and PWN
candidates are reported by \citet{Kargaltsev}, of which 51 with
VHE associations or possible VHE counterparts. The formation of a
pulsar wind is still poorly understood and it is not known which pulsars
are able to drive PWNe and produce VHE radiation. In a PWN, the source of
the energy of the injected electrons is the pulsar spin-down luminosity.
TeV PWNe detected with the current instruments, at a sensitivity of about
2\% of the Crab flux, are mainly associated with young and energetic pulsars
with spin-down power $\dot{E}>10^{35}$ erg s$^{-1}$, showing a TeV luminosity not
significantly correlated with $\dot{E}$ \citep{Kargaltsev, Klepser}.
On the other hand, as mentioned above, several energetic pulsars with
prominent X-ray PWNe are not detected at TeV energies, suggesting that
environmental factors such as, for instance, the local energy density of the
ambient photon field or the intensity of the magnetic field, are relevant
for the evolution of TeV PWNe. The H.E.S.S.
catalog\footnote{http://www.mpi-hd.mpg.de/hfm/HESS/pages/home/sources/},
which reports the results of the Galactic plane survey in the longitude range
$-110^{\circ} < l < 65^{\circ}$, lists 71 TeV sources,
of which about 35\% firmly associated with PWNe and 21\% with
SNRs. Besides a few massive stellar clusters and some binary systems, a large
fraction (about 31\%) remains unidentified, with ambiguous associations or
without any plausible counterpart in X-ray or radio \citep{Carrigan}.
15 TeV sources and candidate sources are recorded in the
$25^{\circ} < l < 65^{\circ}$, $|b|<3.5^{\circ}$ Galactic
region common to H.E.S.S. and ARGO-YBJ. They are associated with PWNe or are
unidentified, but many of them could be old PWNe, still bright in high-energy
gamma-rays, whose synchrotron emission is too faint to be detected in X-rays
\citep{Kargaltsev, Acero}. The ATNF catalog (v1.50,
\citet{Manchester}\footnote{http://www.atnf.csiro.au/research/pulsar/psrcat})
reports 321 radio-loud pulsars in this region. Taking into account the
observational selection effects, the number of predicted pulsars in the
same Galactic region is 4169,
while inside $25^{\circ} < l < 65^{\circ}$, $|b|<5^{\circ}$ is
4633\footnote{with 1.4 GHz luminosities above 0.1 mJy kpc$^2$ and beaming
towards us}. To obtain this prediction, we used the Galactocentric
distribution (model C') suggested by \citet{Lorimer}.
Therefore, the number of TeV PWNe detected at the threshold of 2\% of the Crab
flux is a very tiny fraction of the predicted number of their radio
counterparts. An estimate of the contribution to the measured diffuse emission
from sources with lower fluxes would require a model of a full synthetic
population of PWNe based on H.E.S.S. and \emph{Fermi} data and on the current
theories of PWN evolution (e.g., \citet{Bucciantini, Mayer2012, Torres}).
This study is beyond the scope of the present paper.
However, in order to have an indication of the effect of undetected sources,
we can use the results obtained by \citet{Casanova}.
Taking into account only the sources detected by
H.E.S.S. above 6\% of the Crab flux, these authors determine the
number-intensity relation $dN/dS \propto S^{-2}$, where $S$ is the source
integral flux above 200 GeV, as expected for a uniform density distribution of
sources in a two-dimensional disk. Extrapolating this relation down to 2 mCrab
flux and assuming an average spectral index of -2.3 \citep{Kargaltsev},
we can estimate the cumulative differential flux at 1 TeV as a function of
the number $N$ of unresolved sources with $S$ in the range 0.2\%-2\% of the
Crab flux above 200 GeV \citep{Aharonian2006b}.
As an example, for $N=50$ we obtain a contribution of about 9\% to the diffuse
flux measured by ARGO-YBJ. This contribution depends linearly on $N$, but may
be lower if the source count distribution flattens at low fluxes.
A similar exercise can be applied to SNRs. A study of the contribution of
unresolved shell-type SNRs is carried out by \citet{Volk} to explain the
apparent excess of \emph{Fermi} data at GeV energies in the inner Galaxy. We note
that the number of SNRs predicted in the $25^{\circ} < l < 65^{\circ}$,
$|b|<5^{\circ}$ is 69, using the radial distribution of the surface density
of shell SNRs given in \citet{Case} and the Galactic height distribution given
in \citet{Xu}.

However, these considerations do not settle
definitively this matter. Indeed, the H.E.S.S. survey extends only to
$l=65^{\circ}$ and in a latitude belt $|b|< 3.5^{\circ}$, and the
sensitivity of the instrument is reduced for extended sources. Twenty-two SNRs and
749 PSRs are predicted in the region $65^{\circ} < l < 100^{\circ}$,
$|b|<5^{\circ}$.
Apart from the Cygnus region observed firstly by HEGRA \citep{Aharonian2002}
and then by the Whipple \citep{Konopelko},
MAGIC \citep{Albert2008} and VERITAS \citep{Weinstein, Aliu2013, Aliu2014}
telescopes, this longitude interval has been
surveyed only by Milagro \citep{Atkins04} and ARGO-YBJ \citep{Bartoli2013},
the latter providing a study with a sensitivity of 24\% Crab units.
The H.E.S.S. survey shows clearly a strong decrease of VHE
gamma-ray sources moving toward the outer Galaxy, however the presence of
some isolated sources with fluxes below the ARGO-YBJ sensitivity cannot be
excluded. A single Crab-like source with 10\% of the Crab flux gives a 2\%
contribution at 1 TeV.
The Milagro collaboration has found multi-TeV emission from the direction of
two gamma-ray pulsars (PSR J1928+1746 and PSR J2030+3641) detected by
\emph{Fermi}-LAT \citep{abdo2014}, corresponding to a
total contribution of $(7\pm 2)$\% to the diffuse flux at 1 TeV.
However, for one of these sources only a much lower upper limit has
been obtained by VERITAS, and the flux measured by Milagro might include some
additional diffuse emission.

The \emph{Fermi}-LAT catalog of sources above 10 GeV 1FHL \citep{Ackermann2013}
provides a list of TeV candidates, stating that many of them should be
detectable with the current generation of ground-based instruments, but no
candidate is found lying in the Galactic region
$l=85^{\circ}$ to $l=100^{\circ}$.
In this paper the Fermi Collaboration presents a study of the source
populations above 10 GeV to infer the contribution of the resolved and
unresolved sources to both high-latitude and low-latitude diffuse
backgrounds. The method used in \citet{Strong2007}, based on EGRET data with
predictions for the \emph{Fermi}-LAT, is adopted. Therefore, the source count
distribution for Galactic sources is modeled with
a power law luminosity function $\sim L^{-1.5}$ with given limits, and with a
source distribution in Galactocentric distance based on the model of
\citet{Lorimer} for the pulsar distribution, taken as representative of
Galactic sources. An exponential scale height of 500 pc is assumed. The
simulated differential source count $dN/dS$ is then compared with the
observed flux distribution of the 1FHL sources. Both the source density
and the luminosity range are varied to obtain alternative models. Thus
they estimate that the contribution of sources below the \emph{Fermi}-LAT
detection threshold of $5\times 10^{-10}\; photons\; cm^{-2}\; s^{-1}$ to the
observed gamma-ray intensity above 10 GeV at low latitudes
($|b|<10^{\circ}$, all longitudes), is about 5\%.
This result cannot be easily scaled to the TeV range in the Galactic
region considered here. Nevertheless, it could suggest a small
contribution from unresolved sources. Indeed, for source emission
extending to TeV energies with a spectral index of 2.3, the \emph{Fermi}
threshold flux at 10 GeV corresponds to a flux at 200 GeV about 1.5 times
the threshold used in our study. Moreover pulsars, which are
not expected to contribute to the TeV flux \citep{abdo2013},
account for about half of the Galactic sources used in the \emph{Fermi} estimate.
A similar result has been recently found at 1 GeV considering the
\emph{Fermi}-LAT third source catalog 3FGL \citep{Acero2015}.
In conclusion, we can assume that while the main
contribution from discrete sources has been removed, a residual
contribution from unresolved sources could still affect the measured fluxes.

The ARGO-YBJ data have been used to study the
interval of this region, $40^{\circ} < l < 100^{\circ}$ (section 4.2),
which is not rich in point or extended sources apart from the Cygnus cocoon,
and, in a separate analysis, the innermost part, $65^{\circ} < l < 85^{\circ}$
(section 4.3), which includes the Cygnus region.

\subsection{The Galaxy Region $40^{\circ}< l< 100^{\circ}$,
$|b| < 5^{\circ}$}

\begin{figure}[!t]
\vspace{5mm} \centering \epsscale{.8}
\plotone{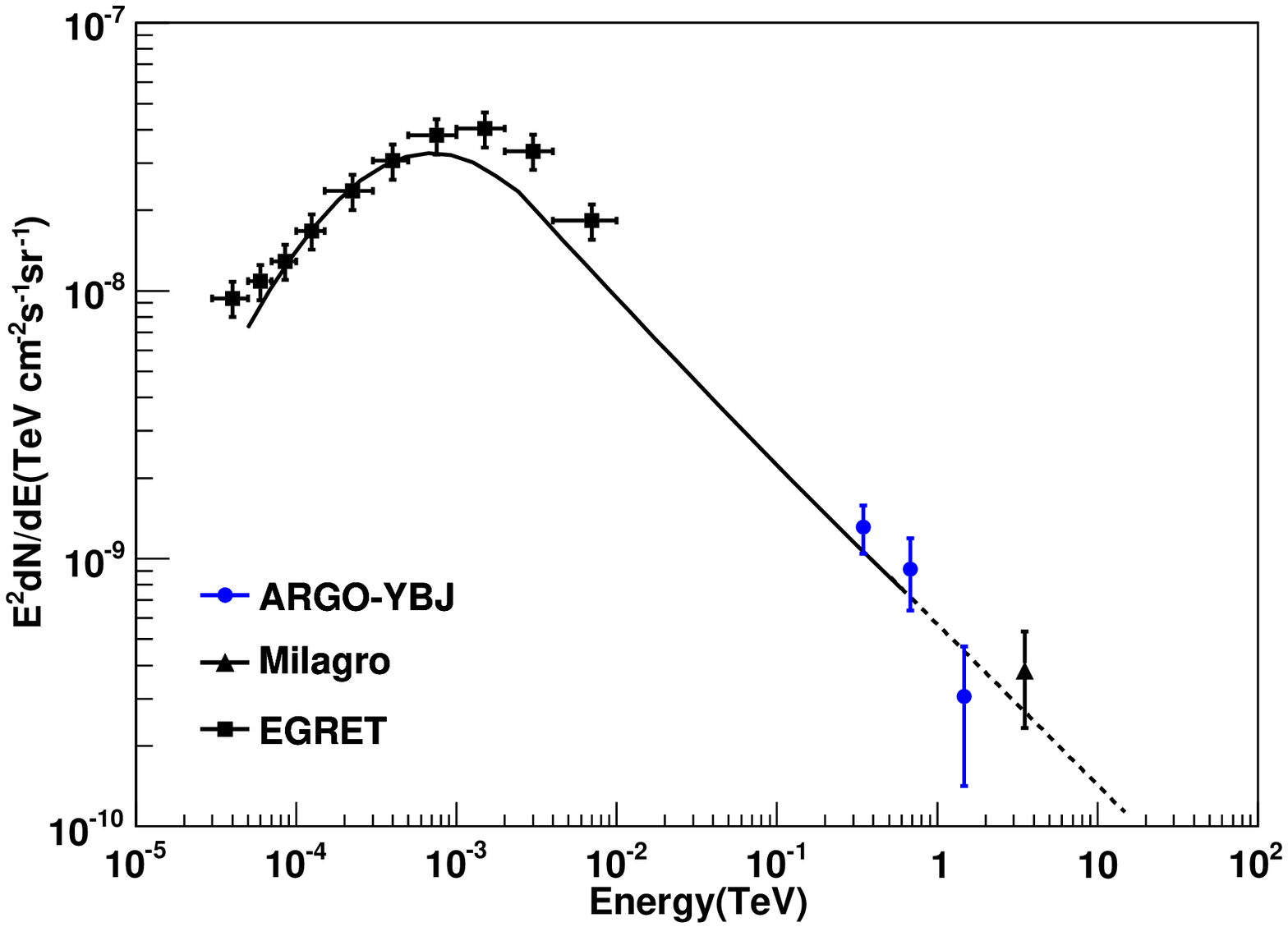}
\caption{Energy spectrum of the diffuse gamma-ray emission measured by
ARGO-YBJ in the Galactic region $40^{\circ}<l<100^{\circ}$, $|b|<5^{\circ}$
(dots). The solid line shows the flux in the same region according to the
\emph{Fermi}-DGE model. The short-dashed line represents its extension following a
power law with spectral index -2.6. The EGRET results (squares) in the same
Galactic region $40^{\circ}<l<100^{\circ}$, $|b|<5^{\circ}$ and the flux measurement by Milagro
(triangle) in the same region are also shown.}
\label{40_100}
\end{figure}

The total number of shower events collected in this region is $7.39\times 10^9$. After
masking the discrete sources and subtracting the residual contribution, an
excess with a statistical significance of 6.1 s.d. above the background is
found. The result of the spectral analysis provides the flux at three median
energies (350 GeV, 680 GeV, and 1.47 TeV, with uncertainties of about 30\%)
as shown in Fig. \ref{40_100} (dots).
The fluxes measured by ARGO-YBJ below 1 TeV are $\sim$20\% larger than
what expected by the \emph{Fermi}-DGE model, but are consistent within the
experimental uncertainties. Fitting the whole ARGO-YBJ data with a power
law we obtain a spectral index of -2.90$\pm$0.31 with a predicted flux at 1
TeV of $(5.2\pm 1.5)\times 10^{-10}$ TeV$^{-1}$ cm$^{-2}$ s$^{-1}$ sr$^{-1}$,
compatible with the extrapolation of the \emph{Fermi}-DGE model.

We have used these data to face the ``TeV excess'' anomaly associated to
the Milagro result concerning this Galactic region.
In fact, in this region the Milagro detector made the first measurement of
the diffuse TeV gamma-ray flux from the Galactic plane \citep{Atkins05}.
The measured flux above 3.5 TeV is $(6.8 \pm 1.5 \pm 2.2)\times
10^{-11}$ cm$^{-2}$ s$^{-1}$ sr$^{-1}$, which, once connected to the EGRET
data with a power law with differential spectral index -2.6, reveals a
``TeV excess'' in the diffuse gamma-ray spectrum, the corresponding flux being
5-10 times higher than expected \citep{Aharonian2008}. In order to
explain the enhanced gamma-ray flux seen by Milagro, other contributions to the
true diffuse flux have been envisaged and  discussed in \citet{TeVexcess},
and include a harder cosmic-ray spectrum,
additional flux from unresolved sources, excess gamma-rays by inverse Compton
scattering and photons from dark matter annihilation. Using the Milagro data,
we converted this integral flux to the differential flux plotted in
Fig. \ref{40_100} (triangle).
This flux is only 34\% greater than the value expected from the \emph{Fermi}-DGE
extrapolation, therefore within the experimental uncertainties. Moreover,
the Milagro result does not take into account the
contributions from the Cygnus cocoon and from the overlapping point or
extended sources TeV J2032+4130, VER J2019+407, and VER J2016+372. Minor
contributions come also from the H.E.S.S. sources. Following the ARGO-YBJ
analysis of the Cygnus cocoon \citep{Bartoli2014a} and taking into account the
width of the latitude band, we can evaluate the fraction of the total flux
generated by these sources. We find that the discrepancy between
the Milagro results and the \emph{Fermi}-DGE predictions is almost canceled out.
The full set of measurements obtained with ground-based experiments is
in agreement with direct observations by \emph{Fermi}-LAT. According to these results
and taking into account that the \emph{Fermi}-LAT data \citep{abdo2009, abdo2010}
do not support the high-intensity diffuse emission observed by EGRET
(the EGRET ``GeV excess''), likely due to instrumental effects
\citep{Stecker}, we rule out the evidence of any ``TeV excess''
requiring additional sources or particle production processes other than
those responsible for the production of Galactic cosmic rays.

\subsection{The Cygnus Region}

\begin{figure}[!t]
\vspace{5mm} \centering \epsscale{.8}
\plotone{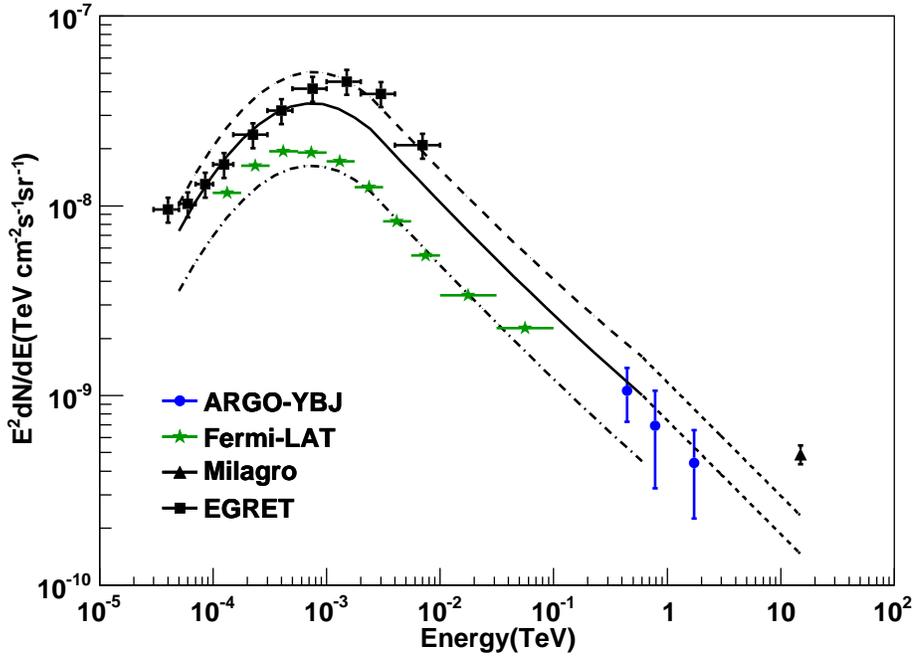}
\caption{Energy spectrum of the diffuse gamma-ray emission measured by
ARGO-YBJ in the Galactic region $65^{\circ}<l<85^{\circ}$, $|b|<5^{\circ}$
(dots). The solid line shows the flux according to the \emph{Fermi}-DGE model.
The short-dashed line represents its extension following a power
law with spectral index -2.6.  The EGRET results (squares) in the same region
are also shown. The Milagro result (triangle) for the Galactic region
$65^{\circ}<l<85^{\circ}$, $|b|<2^{\circ}$ is also given.
The long-dashed line and its extension (short-dashed line)
represent the flux in this region according to the \emph{Fermi}-DGE model.
The spectral energy distribution of gamma-ray emission measured by \emph{Fermi}-LAT
in the Galactic region $72^{\circ}<l<88^{\circ}$, $|b|<15^{\circ}$ is also
reported (stars).
The flux in the same region expected from the \emph{Fermi}-DGE model is shown as a
dot-dashed line.}
\label{65_85}
\end{figure}

The statistical significance of the excess found in the Galactic region
$65^{\circ}< l< 85^{\circ}$, $|b| < 5^{\circ}$ is 6.7 s.d. above the
background. After masking the discrete sources and the Cygnus cocoon and
subtracting the residual contributions, an excess of 4.1 s.d. is left.
This direction points into our spiral arm at the Cygnus star-forming region
hosting a giant molecular cloud complex. Located at a
distance of about 1.4 kpc, this region is rich in potential cosmic-ray
accelerators such as Wolf-Rayet stars, OB associations, and SNRs. Given its
peculiarity, this region has been the target of
numerous multiwavelength observations, including the high-energy
measurements of \emph{Fermi}-LAT (GeV) and Milagro (TeV). \emph{Fermi}-LAT data have
been used to study the region of galactic coordinates
$72^{\circ}< l< 88^{\circ}$, $|b|< 15^{\circ}$, where the bright and extended
cocoon has been observed.
The spectral energy distribution of gamma-ray emission is
shown in Fig. \ref{65_85} (filled stars) and includes different contributions
from the diffuse emission, point sources, and extended objects. A global model
taking into account all the components reproduces satisfactorily the
experimental data and implies that the cosmic-ray flux averaged over the
scale of the whole Cygnus region is similar to that of the local
interstellar space. The expected energy spectrum according to the
\emph{Fermi}-DGE model in the same spatial region is shown for comparison
(dot-dashed line).

This region also received considerable attention by Cherenkov telescopes
and air shower arrays that have discovered VHE emission from point
and extended sources. Recently, the ARGO-YBJ experiment observed the TeV
counterpart of the \emph{Fermi} cocoon \citep{Bartoli2014a}. In this paper a short
summary of the previous observations at TeV energies is also reported.

Exploiting its wide FOV, the Milagro telescope measured the
diffusion emission at Galactic longitudes $65^{\circ}< l< 85^{\circ}$
\citep{abdo2007a}. The first paper reports
the flux in the latitude band $|b|< 3^{\circ}$ at a median
energy of 12 TeV. In a following paper \citep{abdo2008}, more data have been
added and a more refined analysis is applied evaluating the flux at a median
energy of 15 TeV in the region with $65^\circ <l <85^\circ$, $|b|<2^\circ$,
as reported (filled triangle) in Fig. \ref{65_85}. The measured flux is twice
the predictions based on the GALPROP code optimized to reproduce
the EGRET data. This
excess has been attributed to the interaction with the interstellar
medium of hard-spectrum cosmic rays generated by local sources. For
comparison, we show (dashed line) the expected energy spectrum for this
region according to the \emph{Fermi}-DGE model, not available at that time. The
discrepancy is reduced, the Milagro flux being about 75\% higher than the
\emph{Fermi} template, but enough to suggest the presence of an excess.
The flux measured by ARGO-YBJ at median energies 440 GeV, 780 GeV, and 1.73 TeV
(with uncertainties of about 40\%)
and averaged over the latitude band $|b|< 5^{\circ}$ is shown with dots.
The three points can be fitted with a power law with spectral
index -2.65$\pm$0.44. The estimated flux at 1 TeV is
$(6.2\pm 1.8)\times 10^{-10}$ TeV$^{-1}$ cm$^{-2}$ s$^{-1}$ sr$^{-1}$,
resulting about 10\% lower than the \emph{Fermi}-DGE extrapolation.
These data do not show any excess at energies around 1
TeV, corresponding to the excess found by Milagro at an average energy of
15 TeV. One possible explanation of this discrepancy is that the
contribution of all the discrete gamma-ray sources was not completely removed
from the Milagro data. Indeed, the exclusion of discrete sources is of
crucial importance. According to the ARGO-YBJ data, the flux at 1 TeV
injected by the cocoon is of the same order as the diffuse emission flux.

An alternative explanation could be considered if
the spectrum measured in the Cygnus region is compared with that
measured in the complementary part of the $25^\circ <l<100^\circ$ region.
Adding the data from the regions $25^\circ <l<65^\circ$ and
$85^\circ <l<100^\circ$, we found an excess of 5.6 s.d. above the background.
The measured spectrum has an index -2.89$\pm$0.33, while the estimated flux at
1 TeV is $(6.0\pm 1.7)\times 10^{-10}$ TeV$^{-1}$ cm$^{-2}$ s$^{-1}$ sr$^{-1}$.
Thus there is an indication that the spectrum of the diffuse emission
in the Cygnus region could be harder than that in the complementary
part of the $25^\circ <l<100^\circ$ longitude interval.
Assuming that there is actually a difference, a plausible
explanation is that the region of about 500 pc around the Cygnus cocoon is
more abundant of cosmic rays accelerated by a nearby source, which produces
also the TeV emission from the cocoon, whose spectrum has not
yet been steepened by diffusion \citep{Ahatoyan, Gabici}.
These runaway cosmic rays may diffuse to
a characteristic length of a few hundred parsecs \citep{Casanova2010} and
interact with the local gas producing gamma-rays with the same spectral
shape via $\pi^0$ decay. In fact, since the hadronic interactions at
multi-TeV energies are basically scale-invariant and TeV photons are not
attenuated by the interstellar radiation fields \citep{Moskalenko2006},
the gamma-ray spectrum is expected to mimic the cosmic-ray spectrum. This
interpretation assumes a hadronic origin of the gamma-ray emission from
the cocoon as discussed in \citet{Bartoli2014a}.
In this scenario, the region around the
cocoon is expected to contain a mixture of ordinary background cosmic rays
and young cosmic rays with a harder spectrum, released first
from the source, which diffuse fastly and reach a distance depending
upon many factors as the injection history of the source, the diffusion
coefficient and the interstellar gas density. Thus, the diffuse gamma-ray
emission may consist of two distinct components produced by these two cosmic-ray populations which have different spatial extension and different
spectral shape. The superposition of these two components may produce
concave spectra at TeV energies \citep{Gabici}, accounting for
the Milagro result in case there is a residual excess after removing
all the source contributions. An interesting application of these
concepts is given in \citet{Casanova2010}. The ARGO-YBJ data do not allow
the study of this phenomenology. Accurate measurements of the diffuse
gamma-ray emission at TeV energies on an angular scale of a few degrees
are necessary. Future experiments with
higher angular resolution and large FOV, such as HAWC \citep{HAWC}
and LHAASO \citep{Cao}, are expected to probe the spatial distribution of the
photon flux at TeV energies, providing a detailed map of the diffuse
gamma-ray emission in this region. They will also benefit of the CTA
\citep{CTA} survey, expected to detect very faint discrete sources, thus
providing information useful to separate the genuine diffuse emission.

The hadronic origin of the gamma-ray emission from the cocoon can be
also probed searching for an excess of GeV-TeV neutrinos over the atmospheric
neutrino background from the Cygnus cocoon \citep{Tchernin, Niro}.
Interestingly, the energy spectrum of the light
component (protons plus Helium nuclei) of the primary cosmic rays from a
few TeV to 700 TeV measured by ARGO-YBJ \citep{Bartoli2012a} and by the hybrid
experiment ARGO-WFCTA \citep{Bartoli2014b} follows the same spectral shape as
that found in the Cygnus region.
A precise comparison of the spectrum of young cosmic rays, as those supposed in
the Cygnus region, with the spectrum of old cosmic rays resident in other
places of the Galactic plane, could help to determine the distribution of the
sources of the cosmic rays observed at Earth.

\subsection{Outer Galaxy}

\begin{figure}[!t]
\vspace{5mm} \centering \epsscale{.8}
\plotone{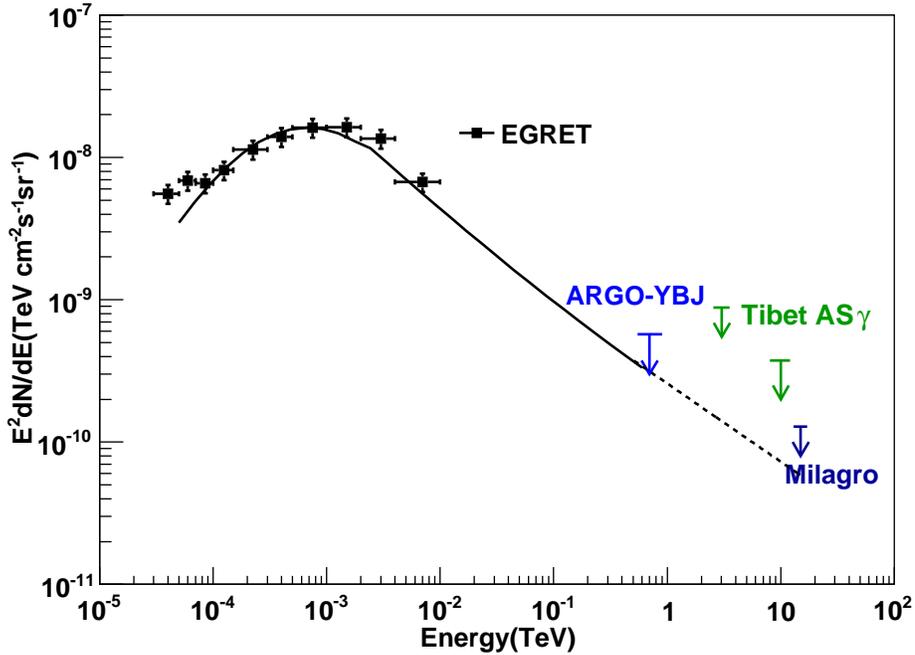}
\caption{The 99\% C.L. upper limit at a median energy of 700 GeV as obtained
by ARGO-YBJ for the Galactic region $130^{\circ}<l<200^{\circ}$,
$|b|<5^{\circ}$. The solid line shows the flux in the same
region according to the \emph{Fermi}-DGE model. The short-dashed line represents its
extension following a power law with spectral index -2.6.  The EGRET results (squares) in the same region
are also shown. For comparison, the upper limits from the Milagro
(95\% C.L., $136^{\circ}<l<216^{\circ}$, $|b|<2^{\circ}$) and
Tibet AS$\gamma$ (99\% C.L., $140^{\circ}<l<225^{\circ}$, $|b|<2^{\circ}$)
experiments are also reported.}
\label{130_200}
\end{figure}

No excess has been measured in the outer Galaxy region
$130^\circ <l<200^\circ$, $|b|<5^\circ$ (after masking the Crab Nebula).
Assuming a spectral index -2.7 the median energy of
all the events with \emph{N}$_{pad} >20$ is 700 GeV. The corresponding upper limit at
99\% confidence level (C.L.) results
$5.7\times 10^{-10}$ TeV$^{-1}$ cm$^{-2}$ s$^{-1}$ sr$^{-1}$ and is
shown in Fig. \ref{130_200}, where the limits
obtained at higher energies by the Tibet AS$\gamma$ (3 and 10 TeV)
and Milagro (15 TeV) experiments
are also reported. The \emph{Fermi}-DGE flux and its extrapolation are shown for
comparison. The ARGO-YBJ upper limit is compatible with the \emph{Fermi} model,
providing an useful constraint to the Galactic diffuse emission around 1 TeV.

\section{Summary and conclusions}

We analyzed the data recorded by ARGO-YBJ over more than five years for a total
live time of 1670.45 days, with the aim of measuring the diffuse gamma-ray
emission at TeV energies in the Galactic region visible from the Northern
Hemisphere. After the application of appropriate selection criteria,
$6.407\times 10^{10}$ high-quality
events are found in the Galactic latitude belt $|b|<15^\circ$.
These events have been used to measure the gamma-ray diffuse emission in the
two Galactic regions $25^\circ <l<100^\circ$, $|b|<5^\circ$ and
$130^\circ <l<200^\circ$, $|b|<5^\circ$ accessible to the
experiment. Fluxes and spectral indexes measured by ARGO-YBJ in these
Galactic regions are reported in Table \ref{table}.
The standard ARGO-YBJ procedure for background subtraction has
been applied, including a suitable approach to correct for cosmic-ray
anisotropy. Great care has been taken in removing the emission from known
gamma-ray sources by masking out the brightest of them and subtracting the
residual contributions. An excess of 6.9 s.d. above the background is
observed in the innermost region $25^\circ <l<100^\circ$, $|b|<5^\circ$,
which has been the target of a detailed
analysis since the pioneering Milagro observations at multi-TeV energies have
shown significant deviations from the predictions based on conventional models
of diffuse gamma-ray emission. As a reference for our results we used
the recent \emph{Fermi} model for diffuse emission extrapolating it to the TeV
region. Firstly, we have studied the region $40^\circ <l<100^\circ$, where a
``TeV excess'' in the diffuse gamma-ray spectrum has been suggested.
The ARGO-YBJ data have been analyzed to derive the differential flux at
three median energies around 1 TeV. Fitting these points with a power law,
we found a spectrum  steeper than the \emph{Fermi}-DGE extrapolation, with
index -2.90$\pm$0.31, however consistent to within 1 s.d..
The large error on the spectral index is due to the short lever arm of these
data and to the poor statistics affecting the highest energy point.
The average flux is compatible with the \emph{Fermi}-DGE extrapolation within
the statistical and systematic uncertainties.
After subtracting the contribution of the gamma-ray sources detected later on
and thus not taken into account, also the flux measured
by Milagro at 3.5 TeV is compatible with the \emph{Fermi}-DGE extrapolation.
Therefore, we cannot confirm the existence of any excess at TeV energies.

\begin{table}[!hbp]
\caption{Diffuse Gamma-ray emission from the Galactic plane for $|b|<5^{\circ}$.}
\begin{tabular}{|c|c|c|c|c|}
  \hline
   l Intervals & Significance  & Spectral Index & Energy (GeV) &Flux$^a$ \\
  \hline
  $25^{\circ}<l<100^{\circ}$ & 6.9 s.d. &  $-2.80\pm 0.26$ & 390 &$8.06\pm1.49$\\\cline{4-5}
  & & & 750&$1.64\pm0.43$\\\cline{4-5}
  & & & 1640&$0.13\pm0.05$\\\cline{4-5}
  & & & 1000$^b$&$0.60\pm0.13$\\\cline{4-5}
  \hline
  $40^{\circ}<l<100^{\circ}$ & 6.1 s.d. & $-2.90\pm 0.31$ & 350 &$10.94\pm2.23$\\\cline{4-5}
  & & & 680&$2.00\pm0.60$\\\cline{4-5}
  & & & 1470&$0.14\pm0.08$\\\cline{4-5}
  & & & 1000$^b$&$0.52\pm0.15$\\\cline{4-5}
  \hline
  $65^{\circ}<l<85^{\circ}$ & 4.1 s.d. & $-2.65\pm 0.44$& 440&$5.38\pm1.70$\\\cline{4-5}
  & & & 780&$1.13\pm0.60$\\\cline{4-5}
  & & & 1730&$0.15\pm0.07$\\\cline{4-5}
  & & & 1000$^b$&$0.62\pm0.18$\\\cline{4-5}
  \hline
  $25^{\circ}<l<65^{\circ}$ \& & 5.6 s.d.  & $-2.89\pm 0.33$ & 380&$9.57\pm2.18$\\\cline{4-5}
  $85^{\circ}<l<100^{\circ}$ & & & 730& $1.96\pm0.59$ \\\cline{4-5}
  & & & 1600&$0.12\pm0.07$\\\cline{4-5}
  & & & 1000$^b$&$0.60\pm0.17$\\\cline{4-5}
  \hline
  $130^{\circ}<l<200^{\circ}$ & -0.5 s.d.  & -- & --& $<5.7^c$\\
  \hline
\end{tabular}

\vspace{1mm}
Note.  The median energies and the corresponding differential fluxes are reported. The errors are only
statistical.

$^a$In units of $10^{-9}$ TeV$^{-1}$ cm$^{-2}$ s$^{-1}$ sr$^{-1}$.

$^b$This entry gives the result of the fit to the three data points.

$^c$99\% C.L. at 700 GeV.
\label{table}
\end{table}

A specific study of the Cygnus region ($65^\circ <l<85^\circ$) is motivated by
the Milagro results showing, at energies $>$10 TeV, a strong enhancement of the
diffuse flux with respect to the model predictions, suggesting the existence
of powerful young accelerators as sources of hard-spectrum cosmic rays.
Indeed, following the \emph{Fermi}-LAT discovery of the Cygnus cocoon at GeV
energies, a TeV counterpart has been reported by ARGO-YBJ, providing an
intense flux of TeV photons. The ARGO-YBJ results on diffuse emission
around 1 TeV do not exhibit any excess when compared to the \emph{Fermi} data at
lower energies, suggesting that the tail of the cocoon flux above 10 TeV
and other contributions from discrete sources not completely removed from
data could explain the excess found by Milagro. The ARGO-YBJ measurements
cover the energy range from about 400 GeV - 2 TeV and follow a power law
with spectral index -2.65$\pm$0.44, a value very close to that found for
TeV emission from the Cygnus cocoon \citep{Bartoli2014a}.
Thus the spectrum appears flatter than
the one found in the whole region $25^\circ <l<100^\circ$
once the Cygnus region is excluded. Indeed, in the combined
region $25^\circ <l<65^\circ$ plus $85^\circ <l<100^\circ$ the spectral
analysis provides an index -2.89$\pm$0.33. These measurements are affected by
large errors and their difference has a marginal statistical significance.

The diffuse gamma-ray flux measured by ARGO-YBJ can provide useful hints
to constrain models of Galactic origin of the high-energy neutrino excess
reported by the IceCube Collaboration \citep{Aartsen}. The origin
of this excess above the atmospheric neutrino background is unknown.
Scenarios invoking an extragalactic origin are favored \citep{ahlzen}.
Alternative models envisaging different possibilities for the
Galactic neutrino sources have been proposed \citep{Ahlers}. One of these
assumes that the TeV-PeV diffuse gamma-ray emission from the Galactic
plane and at least part of the neutrino IceCube excess are produced via
the same mechanism, that is, the interaction of cosmic rays with the
interstellar medium. With this conjecture a clear connection between the
gamma-ray and neutrino fluxes from the Galactic plane can be
established. The TeV-PeV gamma-ray emission should provide a good estimate
for the 100 TeV neutrino signal along the whole Galactic plane. Recent papers
have addressed this scenario \citep{Neronov, Ahlers, Fox}.
The fluxes measured by ARGO-YBJ (see Table 1) add a firm estimate to
the available current data. However, only
upper limits are known for the high-energy diffuse gamma-ray emission (see
Introduction), limiting the capability to draw firm conclusions. Future
gamma-ray observatories as LHAASO \citep{Cao} and HISCORE \citep{hiscore}
will operate with high sensitivity up to PeV energies.
The whole set of TeV-PeV data and an increased statistics of the neutrino
signal from a deeper IceCube exposure are expected to confirm or rule out the
model of a neutrino diffuse emission from the Galactic plane.

In conclusion, the ARGO-YBJ results concerning the diffuse emission at TeV
energies in the $25^\circ <l<100^\circ$, $|b|<5^\circ$ Galactic region
are in agreement with the extrapolation of the \emph{Fermi}-DGE model,
implying that the questions raised by the Milagro observations can be
answered by taking into account the emission of TeV photons from the
Cygnus cocoon and, to a minor extent, from discrete sources.
A spectral analysis of the data has been carried out, showing an energy
spectrum softer than that of the \emph{Fermi}-DGE model, but consistent within
1 s.d.. On the other hand, the TeV flux averaged over the Cygnus region
$65^\circ <l<85^\circ$ shows a marginal evidence of a harder spectrum,
indicating the possible presence of young cosmic rays coming from a nearby
source. Only an upper limit has been set to the diffuse emission in the outer
Galaxy region $130^\circ <l<200^\circ$, $|b|<5^\circ$, but compatible
with the extrapolation of the \emph{Fermi}-DGE model.

\acknowledgments
 This work is supported in China by NSFC (No.10120130794, No.11205165),
the Chinese Ministry of Science and Technology, the
Chinese Academy of Sciences, the Key Laboratory of Particle
Astrophysics, CAS, and in Italy by the Istituto Nazionale di Fisica
Nucleare (INFN).

We also acknowledge the essential supports of W.Y. Chen, G. Yang,
X.F. Yuan, C.Y. Zhao, R. Assiro, B. Biondo, S. Bricola, F. Budano,
A. Corvaglia, B. D'Aquino, R. Esposito, A. Innocente, A. Mangano,
E. Pastori, C. Pinto, E. Reali, F. Taurino and A. Zerbini, in the
installation, debugging and maintenance of the detector.


\clearpage

\clearpage

\end{document}